\documentclass[sigplan,screen]{acmart}
\pdfoutput=1

\usepackage{amsmath}
\usepackage{amsfonts}
\usepackage[ruled,linesnumbered,noend]{algorithm2e}
\usepackage{multicol}
\usepackage{latexsym}
\usepackage{multirow}
\usepackage{graphicx}
\usepackage{pifont}
\usepackage{url}
\usepackage{hyperref}
\newcommand{\ToolName}{\textsc{Falcon}}
\newcommand{\supa}{\textsc{Supa}}
\newcommand{\cred}{\textsc{Cred}}
\newcommand{\dsa}{\textsc{Dsa}}
\newcommand{\sfs}{\textsc{Sfs}}
\newcommand{\saturn}{\textsc{Saturn}}
\newcommand{\compass}{\textsc{Compass}}

\newcommand{\svf}{\textsc{Svf}}

\usepackage{semantic}
\usepackage[utf8]{inputenc}
\usepackage{cleveref}
\crefname{section}{§}{§§}
\Crefname{section}{§}{§§}
\newcommand{\smallletter}[1]{{#1}}

\usepackage{array}
\newcolumntype{L}[1]{>{\raggedright\let\newline\\\arraybackslash\hspace{0pt}}m{#1}}
\newcolumntype{C}[1]{>{\centering\let\newline\\\arraybackslash\hspace{0pt}}m{#1}}
\newcolumntype{R}[1]{>{\raggedleft\let\newline\\\arraybackslash\hspace{0pt}}m{#1}}

\usepackage{balance}

\begin{document}

\title[]{Efficient Path-Sensitive Data-Dependence Analysis}


\author{Peisen Yao}
\affiliation{%
	\institution{The Hong Kong University of Science and Technology}
	\country{Hong Kong, China}
}
\email{pyao@cse.ust.hk}

\author{Jinguo Zhou}
\affiliation{
	\institution{Ant Group}
	\country{China}
}
\email{jinguo.zjg@antfin.com}

\author{Xiao Xiao}
\affiliation{
	\institution{Ant Group}
	\country{China}
}
\email{xx249863@antfin.com}

\author{Qingkai Shi}
\affiliation{
	\institution{The Hong Kong University of Science and Technology}
	\country{China}
}
\email{qingkaishi@gmail.com}

\author{Rongxin Wu}
\affiliation{
	\institution{Xiamen University}
	\country{China}
}
\email{wurongxin@xmu.edu.cn}

\author{Charles Zhang}
\affiliation{%
	\institution{The Hong Kong University of Science and Technology}
	\country{Hong Kong, China}
}
\email{charlesz@cse.ust.hk}

\renewcommand{\shortauthors}{Peisen Yao, Jinguo Zhou, Xiao Xiao, Qingkai Shi, Rongxin Wu, and Charles Zhang}

\begin{abstract}
This paper presents a scalable path- and context-sensitive data-dependence analysis. 
The key is to  address the aliasing-path-explosion problem via
a sparse, demand-driven, and fused approach that piggybacks
the computation of pointer information with the resolution
of data dependence. 
Specifically,
our approach decomposes the computational efforts of disjunctive reasoning into 1) a context- and
semi-path-sensitive analysis that concisely summarizes data
dependence as the symbolic and storeless value-flow
graphs, and 2) a demand-driven phase that resolves transitive
data dependence over the graphs.
We have applied the approach to two clients, namely  thin slicing and value flow analysis. 
Using a suite of 16 programs ranging from 13 KLoC to 8 MLoC,
we compare our techniques against a diverse group of state-of-the-art analyses, illustrating significant precision and scalability advantages of our approach.
\end{abstract}



\maketitle

\section{Introduction}
\label{sec:intro}

Data-dependence analysis aims to identify the def-use information in a
program.  However, the presence of pointers and references obscure such information. The analysis
must cut through the tangle of aliasing to reason about data
dependence, which serves as a substrate for many program
analysis clients, such as impact analysis~\cite{arnold1996software} and program slicing~\cite{sridharan2007thin}.


Path-sensitivity is a common axis for pursuing precision,
yet is stunningly challenging for data-dependence analysis,
which can suffer from the ``aliasing-path-explosion'' problem
For instance, 
at a load statement $p = *x$,  we need to track the path condition of this statement, and the path conditions under which $x$ points-to different memory objects. 
Each load or store statement may access
hundreds of memory objects, each memory object may be accessed at dozens
or hundreds of locations in the program, and the number of calling contexts under which the statements execute can be exponential. 
Consequently, the number of disjunctive cases to track grow extremely large, far too
many to enable a scalable analysis. 


Existing path-sensitive data-dependence analyses can be classified into two major categories. 
The ``fused'' approach can path-sensitively reasons about data-dependence information without a points-to analysis as a priori, such as symbolic execution~\cite{cadar2008klee}.
The fused approach uses various logics 
to generate formulas encoding the entire history of memory writes and reads,  which allow for establishing correlations between variables automatically. 
However, it encodes constraints following control-flow paths, regardless of whether they are relevant to the data dependence of interests or not.
Such ``dense'' analysis is known to have performance problems. 
For instance, \textsc{Focal}~\cite{Focal-FSE19}, a state-of-the-art backward symbolic executor, takes almost 230 hours in answering on-demand queries for a program with near 33 KLoC. 

Alternatively, 
the staged approach
leverages an independent pointer analysis to approximate the def-use information, 
which is then leveraged to bootstrap the path-sensitive data-dependence analysis~\cite{yan2018spatio,blackshear2013thresher}.
Although the idea of leveraging pre-computed pointer information  has advanced flow- and/or context-sensitive analysis (via sparsification~\cite{yan2018spatio}, pruning~\cite{fink:typestate:issta}, or partitioning~\cite{kahlon2008bootstrapping}), how to replicate this success for path-sensitive data-dependence analysis remains an open question.


The problem of computing transitive data-dependence relations
can be formulated as a graph reachability problem, where value-flow graphs varying in precision act as the reachability indices. 
Without an index, the approaches like symbolic execution are hard to scale, whether it be exhaustive or demand-driven.
When computing the index, there is a tension between tracking path-sensitive pointer information too early~\cite{livshits2003tracking,hackett2006aliasing,dillig2011precise}--which leads to the overwhelming cost of value-flow graph construction--and tracking path-sensitive pointer information too late--which reduces the benefits of path-sensitivity, because pointer information can be spuriously and/or redundantly propagated~\cite{yan2018spatio}. 

In this paper, we present \ToolName, a fused and sparse approach 
to path-sensitive data-dependence analysis, which piggybacks the computation of pointer information with the resolution of data dependence.
The key insight is that an data-dependence relation induced by pointer expressions can be identified without knowing the concrete memory objects referenced by the pointers. 
This enables us to efficiently and precisely build a reachability index for value flows, alleviating the need for explicitly and repeatedly enumerating sheer amounts of points-to information.

We first introduce an \textit{all-program-points but lazy} pointer analysis:  it constructs the symbolical and compositional value-flow graphs for the entire program,  without computing exhaustive points-to information. The graphs act as the ``conduits'' for tracking transitive data dependence in a sparse and demand-driven manner. 
Our analysis provides the key precision benefit that path-sensitivity brings, paths pruning and merging, via lightweight semi-decision procedures. 
To achieve context-sensitivity but avoid expensive summary cloning, it only clones the memory access-path expressions that are rooted at a function parameter and incur side-effects,
thus enabling local reasoning of value flows, as opposed to global reasoning about the entire heap.

We then present two client analyses, thin slicing based program understanding~\cite{sridharan2007thin,li2016program} and value-flow analysis based bug hunting~\cite{sui2012static,shi2018pinpoint,yan2018spatio}.
Crucially, the guards qualifying the graph edges have concisely merged value flows going through different memory objects. 
The clients (1) do not need to perform explicit cast-splitting  over the points-to sets when handling indirect reads/writes, thereby alleviating a major source of case explosion in previous path-sensitive analyses~\cite{PSE,blackshear2013thresher,yan2018spatio}, 
because the size of a points-to set could be very large,
and (2) can sparsely track the value flows by following the value-flow edges, aiding scalability.
In summary, our approach separates the task of reasoning about ``how the values flow through different memory objects'' from answering queries about data dependence.

Specifically, there are two novel and critical features in our algorithm itself:
\begin{itemize}
    \item The pointer analysis for building value-flow graphs is both on-the-fly sparse and path-sensitive, in that it computes the heap def-use chains incrementally, along with the path-sensitive pointer information discovered. 
	The \textsc{Spas} algorithm~\cite{sui2011spas} is the only previous work that has the same property, 
	but they achieve incremental sparsity by following the level-by-level analysis~\cite{yu2010level} and, thus, is exhaustive.
    \item When  answering demand data-dependence queries, our analysis can stop as soon as enough evidence is gathered, without trying to find all pointed-by memory objects. 
    The previous analyses~\cite{zheng2008demand,yan2011demand,spath2016boomerang,spath2019context} can answer demand alias queries directly, via different storeless representations~\cite{kanvar2016heap}. 
    However, none of the techniques can introduce path sensitivity.
\end{itemize}

Overall, this paper makes the following key contributions:
\begin{itemize}

	\item 
	We identify and discuss the major challenges  of scaling path-sensitive data-dependence analysis.
		\item We introduce an efficient path-sensitive data-dependence analysis, which, for
		the first time, allow us to analyze multi-million-line
		code bases with the precision of full path-sensitivity
		in minutes.
	\item We demonstrate the utility of our approach with two
	clients, namely thin slicing and value
	flow analysis.
	\item We conduct a significant experiment on 16 real-world programs ranging from 13 KLoC to 8 MLoC.
	\begin{itemize}
		\item 	In building value-flow graphs, \ToolName\ outperforms  \svf~\cite{sui2016svf}, \sfs~\cite{hardekopf2011flow}, and \dsa~\cite{lattner2007making}, achieving on average 17$\times$, $25\times$, and 4.4$\times$ speedups, respectively.
		\item Compared with \supa~\cite{sui2016demand,sui2018vfdemand}, the state-of-the-art demand-driven flow- and context-sensitive pointer analysis for C/C++, \ToolName\ is  54$\times$ in
		answering thin slicing queries,
		 and   it improves the precision by 1.6$\times$.
		\item In comparison with \cred~\cite{yan2018spatio}, a state-of-the-art path-sensitive value flow analysis for bug hunting, \ToolName\ is on average 6$\times$ faster, and finds more real bugs (21 vs. 12) with a lower false-positive rate (25\% vs. 47.8\%).
	\end{itemize}
\end{itemize}

\section{Overview}
\label{sec:motivation}

\begin{figure*}[t]
	\centering
	\includegraphics[width=0.82\textwidth]{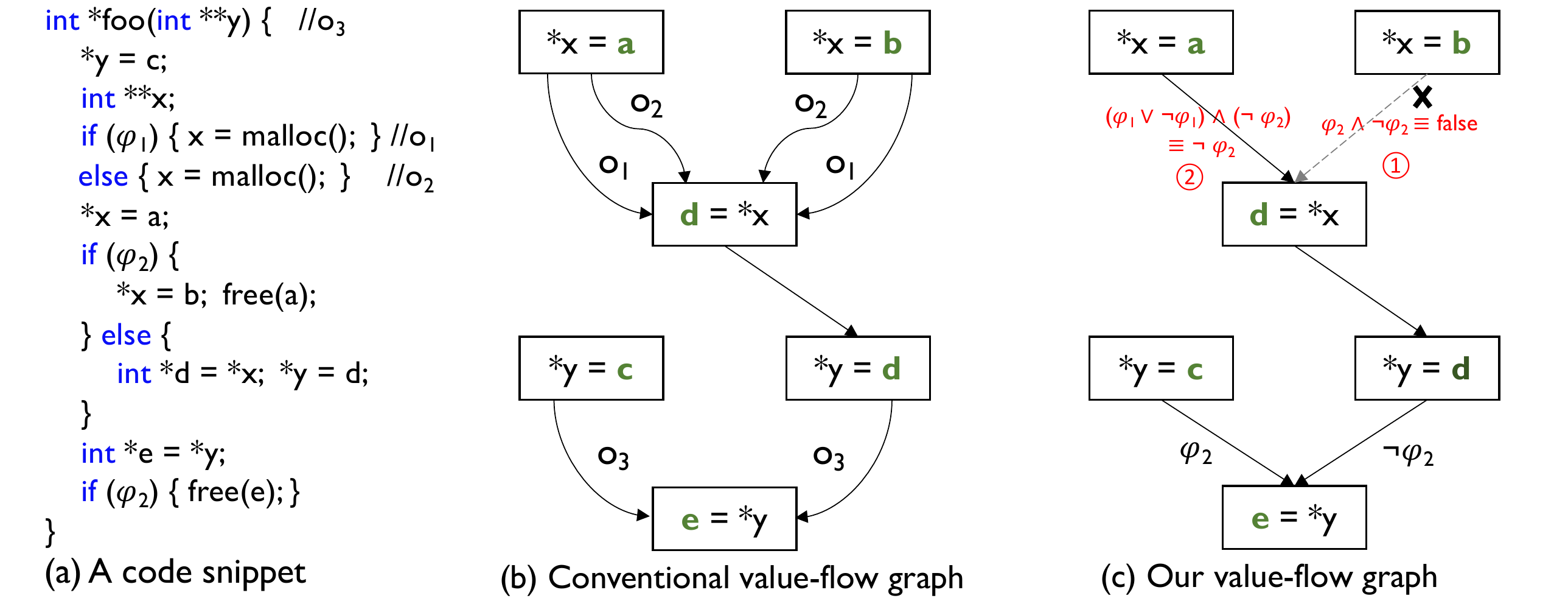}
	\caption{
		An example comparing the conventional value-flow graph and our value-flow graph for answering demand queries. In (b), the label on an edge represents a memory object. In (c), the label on an edge is a path condition.}
	\label{demand-merge}
\end{figure*}

We take the program in Fig.~\ref{demand-merge} as an example to motivate the path-sensitive data-dependence analysis, highlight its challenges, and explain the essence of our approach.

\paragraph{\textbf{Importance of Path-Sensitive Data-Dependence  Information}}
Suppose we need to detect double-free bugs for the program in Fig.~\ref{demand-merge}(a).  
In this program, 
there are two memory deallocation statements \textit{free(a)} and \textit{free(e)}.
Observe that the value of $a$ can flow to $e$ only under the condition $\neg \phi_2$.
Thus, the program is safe, because the deallocation statements execute under the condition $\phi_2$.

Assume that we only approximate the data-dependence information with a path-insensitive pointer analysis, and then partially track path correlations of the memory deallocation statements. 
The pointer analysis can tell that $e$ may be data-dependent on $a$, meaning that $e$ and $a$ may point-to the same memory object.
Observe that the path conditions for the two statements 
\textit{free(a)} and \textit{free(e)}  are both $\varphi_2$, which do not conflict with each other. 
Here, if taking $\varphi_2$ as the path condition of a double-free vulnerability, our analysis would raise a false alarm, because the condition for $e$ to be data-dependent on $a$ is $\neg \varphi_2$. 
To summarize, the imprecise data-dependence information caused by the points-to analysis would be passed on to the clients, hurting the precision. 

\paragraph{\textbf{Problem of Aliasing-Path-Explosion}}
However, obtaining path-sensitive data-dependence information is far from trivial, because tracking path-sensitive pointer information can require reasoning about a considerable number of disjunctive cases.
For instance, 
at a load statement $p = *x$,  we need to track the path condition of this statement, and the path conditions under which $x$ points-to different memory objects. 
each load or store statement may access
hundreds of memory objects, each memory object may be accessed at dozens
or hundreds of locations in the program, and the number of calling contexts under which the statements execute can be exponential. 
In summary,  
the transfer function for 
each statement needs to store and propagate an enormous amount of information.
We term the problem \emph{aliasing-path-explosion}: analyzing path-sensitive data-dependence information can lead to reasoning about an excessive number of paths~\cite{blackshear2013thresher}.

\paragraph{\textbf{The State-of-the-Art}}
A recent analysis~\cite{yan2018spatio} has leveraged the idea of \emph{sparsity} to refine the flow-insensitive results into a path-sensitive one on demand.
It first constructs the flow-insensitive def-use chains with a pre-analysis,
which then enable the primary path-sensitive analysis to be performed sparsely~\cite{sui2016demand,sui2018vfdemand,yan2018spatio}.
For instance, as shown in Fig.~\ref{demand-merge}(b), the two edges between $*x = a$ and $d = *x$ state that the value of pointer
$a$ can flow to the pointer $d$ via the memory objects $o_1$ or $o_2$, implying that $d$ may be data-dependent on $a$. 
The pre-computed def-use chains enable the primary path-sensitive analysis to be performed sparsely~\cite{sui2016demand,sui2018vfdemand,yan2018spatio}.

However, the flow-insensitive pre-analysis drops path information.
When answering demand queries, the primary analysis still suffers from the aliasing-path-explosion. For example, suppose a client asks, ``what are the set of variables $e$ may be data-dependent on?'' 
Following their work~\cite{yan2018spatio}, we perform an on-demand backward traversal from $e = *y$ to $*x = a, *x = b, *y = c$, respectively. Apparently, in the worst case, such graph traversal needs to search five paths and solve five path constraints. 
This number of paths exceeds the total number of paths in the program (which is four), 
meaning that the aliasing-path-explosion can be even worse than the well-known scalability problem caused by conditional branching in symbolic execution. 

In essence
existing sparse analysis can use a pre-analysis to identify the relevant memory objects (as in Fig.~\ref{demand-merge}(b)).
However, 
\textit{the pre-analysis  can only reduce the number of  memory objects to track, but not the number of value flow paths going through the relevant memory objects, because it is unaware of the path conditions qualifying the value flows}.
When going for path sensitivity, the primary path-sensitive analysis phase 
cannot avoid aliasing-path-explosion.


\paragraph{\textbf{Our Approach}}
At a high level, our approach works in two phases. 
In the first phase, we compute the  guarded and storless value-flow graphs. 
In the second stage, the clients can utilize
the graphs to track transitive program dependence on demand.
The crux is to judiciously merge abstract states while building the graphs, since when and how to merge them drastically affect the  accuracy and performance of the client analyses.

We observe that many memory objects and paths qualifying data-dependence relations are redundant, which can be symbolically identified and merged. For example, intuitively, the variable $d$ may be data-dependent $a$, no matter $x$ points-to $o_1$ or $o_2$. 
However, the analysis in the previous work~\cite{yan2018spatio} \emph{has to} separate the two edges and label them with different memory objects, so that it can preserve the capability of precision refinement based on the memory objects.

Based on the observation, our key idea is to use a symbolical storeless representation for pointer expressions, which concisely ``index'' how values flow in and out of the memory in a precision-preserving manner.
Crucially, as illustrated in Fig.~\ref{demand-merge}(c), our first phase takes advantage of a lightweight semi-decision procedure to achieve the following two merits,
which significantly reduces the burden of the second phase:
\ding{172} it can efficiently prune a number of infeasible value flows;
and 
\ding{173} it can effectively merge and simplify path constraints at the time of merging value-flow edges. 
Besides, to build interprocedural value-flow graphs, it only clones the memory access-path expressions that are rooted at a function parameter and incur side-effects, there by avoiding computing a whole-program image of the heap.

Then,
in the second phase, we can answer demand data-dependence queries for variables of interest.
For example, 
consider the graph in Fig.~\ref{demand-merge}(c), where the memory objects pointed-by $x$ and $y$ are implicit.

\begin{itemize}
	\item To determine the values that  $e$ is data-dependent on, we perform a backward graph traversal, and only have to traverse  two paths, i.e., from $e$ to $c$ and $a$, respectively. 
	\item To detect double free bugs, we perform a forward traverse from $a$ to $e$, collecting the guard qualifying the edges (i.e., $\neg \varphi_2 \land \neg \varphi_2$). We then collect the  path conditions of the two statements \textit{free(a)} and \textit{free(e)} (i.e., $\varphi_2 \land \varphi_2$).
	Clearly, we can eliminate the false positive because $\neg \varphi_2 \land \varphi_2$ is unsatisfiable.
\end{itemize}

\section{Preliminaries}
\label{sec:domain}
This section presents the basic terminologies and notations used in the paper,
including the language, abstract domains, as well as 
the guarded value-flow graph.

We formalize our analysis with a simple language, as  in Fig.~\ref{lang-syntax}. Programs are in the static single assignment form.
Statements include address-taken statements, common assignments, $\phi$-assignments, loads, stores, branches, returns, procedure calls, and sequencing. 
In the $\phi$-assignment, $\gamma_i$ is the gated function for each $v_i$, which means $v=v_i$ if and only if $\gamma_i$ is satisfied.
Such gated functions can be computed in almost linear time~\cite{ottenstein1990program}.
With no loss of generality, we assume each function has only one return statement.

\begin{figure}[t]
	\begin{align*}
	\begin{split}
	\textit{Program} \ \ P    &:=  F+ \\
	Function \ \ F   &:= f(v_1, v_2, ...) \{ S;\}  \\
	Statement \ \ S   &:= v_1 = \&v_2  \\ 
	& \ | \  v_1 = v_2  \ | \ v = \phi((\gamma_1, v_1), (\gamma_2, v_2),...) \\
	& \ | \  v_1 = *v_2   \ | \  *v_1 = v_2 \\
	& \ | \ \textit{if} \ (v) \ \textit{then} \ S_1 \ \textit{else} \ S_2 \ | \ \textit{return} \ v \\
	& \ | \ r = \textit{call} \ f(v_1, v_2, ...) \ | \ S_1; S_2 
	\end{split}
	\end{align*}
 	\caption{
 	The syntax of the language.}
    \label{lang-syntax}
\end{figure}

\paragraph{\textbf{Abstract Domains}}
The symbols and abstract domains are listed in Fig.~\ref{lang-domain}. 
A label $\ell \in \mathcal{L}$ indicates the position of a statement in the control flow graph.
An {abstract value} $v \in \mathcal{V}$ is a pointer that points-to a memory location. 
A {memory object} $o \in \mathcal{O}$ represents a memory location that may contain different abstract values on different guard conditions $\psi$.
We {factor} the abstract domain to the points-to environment $\mathbb{E}$ and abstract store $\mathbb{S}$.
$\mathbb{E}(v) = \{(\psi, o)\}$ means that the pointer $v$ points-to the memory object $o$ under the condition $\psi$.
$\mathbb{S}(o) = \{(\psi, \ell, v)\}$ states that the memory object contains the value $v$, which is stored into the memory object at the program point $\ell$ on the condition $\psi$.
For simplicity, we define the operation $\Pi_{\psi}$, so that we can query $\mathbb{E}$ and $\mathbb{S}$ under a precondition $\psi$.
Formally,
$$\Pi_{\psi}(\mathbb{S}(o)) = \{ (\pi \land \psi, \ell, v)  \ | \  (\pi,  \ell, v) \in \mathbb{S}(o)  \}$$
$$\Pi_{\psi}(\mathbb{E}(v)) =  \{ (\pi \land \psi, o)  \ | \  (\pi, o) \in \mathbb{E}(v)  \}$$    

\begin{figure}[t]
	\begin{align*}
	\begin{split}
	Labels  \ \ \ell & \in \mathcal{L} \\
	Values  \ \ v & \in \mathcal{V} \\
	Objects \ \ o & \in \mathcal{O} \\ 
	Guard \ \ \psi &:= \textit{true} \   |   \  \textit{false} \   |  \  \psi_0  \land \psi_1 \ | \ \psi_0 \lor \psi_1 \ | \ \neg \psi \\
	Environment  \ \ \mathbb{E} &:= \mathcal{V}  \to  2^{(\psi, \mathcal{O} )} \\
	Store \ \ \mathbb{S} &:= \mathcal{O} \to  2^{(\psi, \mathcal{L}, \mathcal{V})}  \\
	\end{split}
	\end{align*}
	\caption{
		The abstract domains.}
	\label{lang-domain}
\end{figure}

\paragraph{\textbf{Guarded Value-Flow Graph}}
Intuitively, a value 
$q$ flows to $p$ if $q$ is assigned to $p$ directly (via an assignment, such as $p=q$) or indirectly (via pointer dereferences, such as  $*x = q;y = x; p = *y;$). 
Formally, we define the guarded value-flow graph as below.
\begin{definition} (Guarded Value-Flow Graph)
	\label{def:gvfg}
	A guarded value-flow graph is a directed graph $\mathcal{G} = (\mathcal{N}, \mathcal{E}, \mathcal{C})$, where $\mathcal{N}$, 
	$\mathcal{E}$, and $\mathcal{C}$ are defined as following:
	\begin{itemize}
		\item $\mathcal{N}$ is a set of nodes, each of which is denoted by $v \smallletter{@} \ell$, meaning that the variable $v$ is defined or
		used at a program location $\ell$.
		\item $\mathcal{E} \subseteq \mathcal{V} \times \mathcal{V}$ is a set of edges, each of which represents a value-flow relation.
		$(v_1 \smallletter{@} \ell_1, v_2 \smallletter{@} \ell_2) \in \mathcal{E}$ means that the value $v_1 \smallletter{@} \ell_1$ flows to $v_2 \smallletter{@} \ell_2$.
		\item $\mathcal{C}$ maps each edge in the graph to a condition $\psi$, meaning that the value-flow relation holds only when the condition is satisfied.
	\end{itemize}
\end{definition}
Our approach first computes a guarded value-flow graph for each function. A specific client of data-dependence analysis can then be reduced to graph reachability problems, whereby the local value-flow graphs are stitched together by matching formal and actual parameters as well as return value and its receivers. 



To achieve path-sensitivity, for a value-flow edge that only holds under some condition, we label the edge with the constraint $\psi$. 
To establish such guarded edges, 
the key is to record what values will be stored into a memory object at a store statement (e.g.,~$*x = q$) and
query what values can be loaded at a load statement (e.g.,~$p = *y$).
We will detail this process in the following sections. 

%

%

\section{Intraprocedural Analysis}
\label{sec:intra}	
This section presents our intraprocedural analysis to compute $\mathbb{E}$ and $\mathbb{S}$, so that we can establish indirect value flows by querying what values can be loaded at a load statement.
We first define the abstract transformers, which enables a conventional data-flow analysis.
At the end of~\cref{subsec:rules}, we summarize the challenges for optimization, which are addressed in ~\cref{subsec:sparse} and~\cref{subsec:pathupdate}.

\subsection{Abstract Transformers}
\label{subsec:rules}

Fig.~\ref{intra-rules} lists the rules for analyzing the basic statements.
The rule for $\ell, \psi:stmt$ states that under the current points-to environment $\mathbb{E}$, abstract store $\mathbb{S}$, and path condition $\psi$, the statement $stmt$ produces new points-to environment $\mathbb{E'}$ and/or abstract store $\mathbb{S'}$.

\begin{figure}[t]

	\[
	\inference[\textsc{addr} ($\ell, \psi : p = \&a$)]
	{
		\text{ }
	}
	{
		\mathbb{E}'(p) = (\psi, \textit{alloc}_a) 
	}
	\]
	
	\[
	\inference[\textsc{copy} ($\ell, \psi : p = q$)]
	{
		\forall(\pi, o) \in \Pi_{\psi}(\mathbb{E}(q))
	}
	{
		\mathbb{E}'(p) = \mathbb{E}(p) \cup \{ (\pi, o) \}
	}
	\]
	
	\[
	\inference[\textsc{phi} ($\ell, \psi : p = \phi((\gamma_1, p_1),..., (\gamma_n, p_n)$)]
	{
		\forall i \in [1, n],  \\
		\forall(\pi, o) \in \Pi_{\gamma_i}(\mathbb{E}(p_i))
	}
	{
		\mathbb{E}'(p) = \mathbb{E}(p) \cup \{ (\pi, o) \}
	}
	\]
	
	\[
	\inference[\textsc{store} ($\ell, \psi : *x = q$)] 
	{
		\forall(\pi, o) \in \Pi_{\psi}(\mathbb{E}(x)) \\
	}
	{
		\mathbb{S'}(o) = \mathbb{S}(o) \cup \{ (\pi, \ell, q) \} 
	}
	\]
	
	\[
	\inference[\textsc{load} ($\ell, \psi : p = *y$)]
	{
		\forall(\pi, o) \in \Pi_{\psi}(\mathbb{E}(y)) \\
		\forall(\varphi, \ell', v) \in \Pi_{\pi}(\mathbb{S}(o)) \\
	}
	{
		\mathbb{E'}(p) = \mathbb{E}(p) \cup \Pi_{\varphi}(\mathbb{E}(v))
	}
	\]
	\par
   
	\caption{Basic rules for updating $\mathbb{E}$ and $\mathbb{S}$.}	
    \label{intra-rules}
\end{figure}

Rule \textsc{addr} creates memory objects at allocation sites. 
Rule \textsc{copy} and Rule \textsc{phi} are self-explanatory; thus,
we focus on the \textsc{store} and \textsc{load} rules.

Rule \textsc{store} processes a store statement $*x = q$ under path condition $\psi$,
which results in new configurations of the abstract store $\mathbb{S}$.
We first query the memory objects $x$ may point-to, denoted $\Pi_{\psi}(\mathbb{E}(x))$.
For all guarded memory objects $(\pi, o) \in \Pi_{\psi}(\mathbb{E}(x))$, we update the abstract store $\mathbb{S}$ to record the values that $o$ may hold. Following conventional singleton-based algorithms,  if $x$ points-to at most one concrete memory object, we can perform an indirect strong update, which kills other values hold by the memory object $o$~\cite{emami1994context,hardekopf2009semi,hardekopf2011flow,lhotak2011points,yu2010level}.

Given a load statement $p = *y$ under path condition $\psi$ at program location $\ell$, we apply Rule \textsc{load} as follows.  Similar to the \textsc{store} rule, we query the memory objects that $y$ may point-to under the condition $\psi$, denoted  $\Pi_{\psi}(\mathbb{E}(y))$.  
We then fetch the values from each memory object $(\pi, o) \in \Pi_{\psi}(\mathbb{E}(y))$, denoted $\Pi_{\pi}(\mathbb{S}(o))$.   
Finally, for every $(\varphi, \ell', v) \in \Pi_{\pi}(\mathbb{S}(o))$,  we update $\mathbb{E}$ by adding the points-to set of $v$ under condition $\varphi$ as a subset of the points-to set of $p$.

\paragraph{\textbf{Merging Value-Flow Edges}}
Recall from~\cref{sec:domain} that our analysis computes a guarded value-flow graph that summarizes value flows induced by the memory. 
We formalize the rule of building indirect value-flow edges in Fig.~\ref{vfrule}. 
The \textsc{vflow} rule states that when $q \smallletter{@} \ell_1$ and $p \smallletter{@} \ell_2$ are stored and loaded from 
the same memory object $o$, 
$p \smallletter{@} \ell_2$ may alias with $q \smallletter{@} \ell_1$. 

\begin{figure*}[t]
	\centering
	\[
	\inference[\textsc{ \qquad \qquad vflow}($\ell_1, \psi_1: *x = q; \ \cdots; \   \ell_2, \psi_2 : p = *y$)]
	{
		\forall (\pi_i, o_i) \in \Pi_{\psi_2}(\mathbb{E}(y)) \\
		\forall (\varphi_i, \ell_1, q) \in \Pi_{\pi_i}(\mathbb{S}(o_i))
	}
	{ (q@\ell_1, p @\ell_2) \in \mathcal{E}, \mathcal{C}((q@\ell_1, p @\ell_2)) = \vee\varphi_i}
	\]
	
	\caption{Building indirect  value-flow edges.}	
	\label{vfrule}
\end{figure*}

In the rule,
suppose $\Pi_{\psi_2}(\mathbb{E}(y))=\{ (\pi_1, o_1), (\pi_2, o_2) \}$
such that $(\varphi_1, \ell_1, q) \in \Pi_{\pi_1}(\mathbb{S}(o_1))$
and $(\varphi_2, \ell_1, q) \in \Pi_{\pi_2}(\mathbb{S}(o_2))$.
As illustrated in Fig.~\ref{demand-merge}(b),
conventional approaches build flow-insensitive value-flow edges~\cite{sui2016demand,yan2018spatio}. 
Thus, they have to distinguish a value flow induced by different memory objects,
so that they can preserve the capability of precision refinement based on the memory objects.
Such methods, however, can still suffer from the aliasing-explosion-problem, making the analysis not scalable. 

To tackle the problem, the \textsc{vflow} rule merges  the value-flow edges, which not only reduces the number of edges, but also can normalize and simplify the conditions based on some simple rewriting rules.
For example, when merging two value-flow edges under the conditions $\psi \land \pi$ and $\neg \psi \land \pi$ respectively,
the  condition can be simplified as $\pi$ after merging.

\paragraph{\textbf{Challenges}}
However, a 
highly precise (e.g., flow- and path-sensitive)
analysis that uses the above rules to compute value-flow edges is notoriously expensive, due to the following challenges:
\begin{enumerate}
\item \label{challenge1} \emph{Conservative propagation}.
Propagating data-flow facts along control flows is expensive and unnecessary~\cite{hardekopf2011flow}.
To mitigate this problem, data-flow facts can be propagated along with def-use chains. 
However, the def-use information of memory objects is unavailable without a pointer analysis.
To resolve the paradox, most existing work~\cite{hardekopf2011flow,ye2014region,sui2016sparse,sui2016demand,yan2018spatio} perform a lightweight but imprecise pointer analysis to over-approximate the def-use chains. Due to the imprecision, many false def-use relations are introduced, hurting performance.

\item \label{challenge2} \emph{Constraints explosion}.
Our analysis needs to account for a sheer number of guard updates for each statement, quickly causing the explosion of constraints. For a demand-driven analysis, it is both intractable and unnecessary to pay the full price of path-sensitive reasoning up front.
\end{enumerate}

We address these challenges by the following means:
\begin{itemize}
\item The first challenge is addressed via an \emph{on-the-fly sparse} analysis,  which computes def-use relations incrementally during the analysis, instead of relying on precomputed imprecise def-use chains (\cref{subsec:sparse}).

\item The second challenge is addressed via a \emph{semi-path-sensitive} analysis that simplifies and partially solves the constraints, to merge redundant value-flow edges and  prune obviously false value-flow edges (\cref{subsec:pathupdate}).
\end{itemize}

\subsection{On-the-Fly Sparse Analysis}
\label{subsec:sparse}
To address the challenge of conservative propagation, we utilize the idea of sparsity to skip unnecessary control flows when propagating data-flow facts. To this end, instead of leveraging the imprecise def-use relations computed by a pre-analysis, we construct the def-use relations incrementally during the analysis, along with the precise pointer information discovered. 
To formally present the idea, we maintain the abstract store $\mathbb{S}$ as a set of $\mathbb{S}_{\mathsf{b}}$,
which describes the abstract store at the basic block $\mathsf{b}$.  
Then the \textsc{store} and \textsc{load} rules are refined as follows.

\paragraph{\textbf{The \textsc{store} Rule}}
We follow the idea in SSA form where a variable defined at a basic block $\mathsf{b}$ can only be used in a basic block dominated by $\mathsf{b}$ or in the dominance frontier where the definition is merged with other definitions~\cite{cytron1991efficiently}.
Suppose at basic block $\mathsf{b}$, a store statement writes a guarded value $(\varphi,  \ell, v)$ to the memory object $o$.  
As shown in Alg.~~\ref{propagate-def}, it takes two steps to update the abstract store. 
First, we write  $(\varphi, \ell, v)$ into the local store $\mathbb{S}_{\mathsf{b}}(o)$ (Line 3).
Second, we propagate the abstract value to the dominance frontiers of $\mathsf{b}$. We update the guard for the propagated abstract value, which is the conjunction of $\varphi$ and the path condition of $\mathsf{b}$ (Lines 4-5).
Note that it is unnecessary to propagate the abstract value to basic blocks dominated by $\mathsf{b}$, because at the load time, we can walk up the dominance tree to find the corresponding definitions (see the next paragraph).

\begin{algorithm}[t]
	\caption{Write a value to memory objects and propagate the value}
	\label{propagate-def}
	\KwIn{A store statement $\ell, \psi, *x = v$}
	\KwOut{Update the abstract store $\mathbb{S}$}
	$\mathsf{b} \leftarrow$ the basic block of $\ell$\;
    \For{$(\pi, o) \in \Pi_{\psi}(\mathbb{E}(x))$} {
    	$\mathbb{S}_{\mathsf{b}}(o) \leftarrow \mathbb{S}_{\mathsf{b}}(o) \cup \{ (\pi, \ell, v) \}$ \;
    	\For{$\mathsf{b}'$ \text{is a dominance frontier of } $\mathsf{b}$}{ 
    		$\mathbb{S}_{\mathsf{b}'}(o) \leftarrow \mathbb{S}_{\mathsf{b}'}(o) \cup \{ (\pi \land \psi, \ell, v) \}$ \;
    	} 
    }
\end{algorithm}


\begin{figure}[t]
	\centering
	\includegraphics[width=0.48\textwidth]{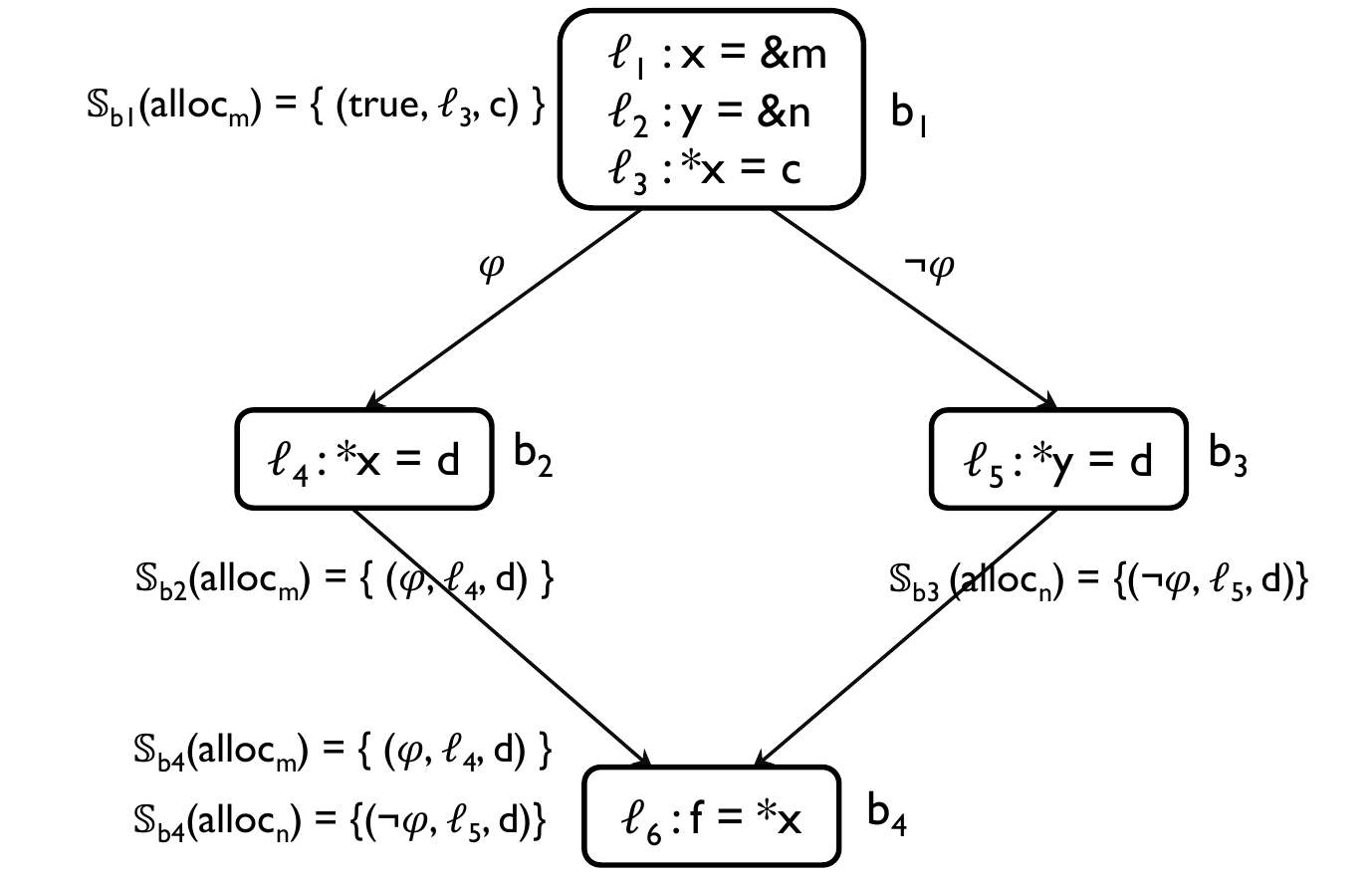}
	\caption{
		An example on sparse analysis.}
	\label{sparse}
\end{figure}

\begin{example} \label{exmp:storevalue}
	Consider the program in Fig.~\ref{sparse}. After $\ell_1, \ell_2$, the variables $x, y$ points-to \textit{alloc}$_m$, \textit{alloc}$_n$, respectively. 
	Suppose we are analyzing the store statement $*x = d$ at basic block $\mathsf{b}_2$. The abstract value $(\psi, \ell_4,  d)$ is stored into  \textit{alloc}$_m$  and then propagated to the local abstract store of $\mathsf{b}_2$'s dominance frontier  $\mathsf{b}_4$. Therefore, we have 	 $\mathbb{S}_{\mathsf{b}_2}(\textit{alloc}_m) = \{ (\psi, \ell_4,  d) \}$ and $\mathbb{S}_{\mathsf{b}_4}(\textit{alloc}_m) = \{(\psi, \ell_4,  d)\}$. Similarly, after analyzing the store statement  $*y = d$, we have $\mathbb{S}_{\mathsf{b}_3}(\textit{alloc}_n) = \{ (\neg \psi, \ell_5,  d) \}$ and $\mathbb{S}_{\mathsf{b}_4}(\textit{alloc}_n) = \{ (\neg \psi, \ell_5,  d) \}$.
\end{example}

\paragraph{\textbf{The \textsc{load} Rule}}
As shown in Alg.~~\ref{getvalues}, for a load statement $u = *x$ at the basic block $\mathsf{b}$, we track the values that can be read from the memory objects pointed-by $x$. 
For each memory object $o$, 
it suffices to walk up the dominance tree (Lines 6-15) to gather abstract values until a strong update is found.
The basic idea behind the approach is that the definition of a variable must dominate its uses.
Clearly, this is a linear search in the dominance tree.

\begin{algorithm}[t]
	\caption{ Read values from memory objects by walking up the dominator tree}
	\label{getvalues}
	\KwIn{A load statement $\ell, \psi: u = *x$}
	\KwOut{Values that can be loaded from $*x$}
	\SetKwFunction{FMain}{LoadInst}
	\SetKwFunction{ReadFromObject}{ReadFromObject}
	\SetKwProg{Fn}{Function}{:}{}
	
	$R  \leftarrow \emptyset$; $\mathsf{b}  \leftarrow$ the basic block of $\ell$\;
	\For{$(\beta, o) \in \Pi_{\psi}(\mathbb{E}(x))$}{
		$R \leftarrow R \ \cup$ \texttt{ReadFromObject}($\beta, o, \mathsf{b}$)\;
	}
	\Return $R$\;
	\SetKwProg{Pn}{Function}{:}{\KwRet}
	\Pn{\ReadFromObject{$\beta, o, \mathsf{b}$}}{
		$\sigma  \leftarrow \texttt{true}, R_o \leftarrow \emptyset$\;
		\While{$\mathsf{b} \neq null$} {
			\For{$(\pi, \ell', v) \in \mathbb{S}_{\mathsf{b}} (o)$}{
				$\varphi \leftarrow \pi \land \sigma \land \beta$\;
				\If{$\varphi$ is satisfiable}{ 
					$R_o \leftarrow R_o \cup \{ (\varphi, \ell', v) \}$\;
				}
				\If{$(\pi, \ell', v)$ is a strong update at $\mathsf{b}$}{ \Return $R_o$;}
				$\sigma \leftarrow \neg \pi \land \sigma$ \;
			}
			$\mathsf{b} \leftarrow$ the immediate dominator of $\mathsf{b}$;
		}
		\Return $R_o$\;
	}
\end{algorithm}

\begin{example} \label{exmp:loadvalue}
	Consider the program in Fig.~\ref{sparse}. When analyzing the load statement $f = *x$ at the basic block $\mathsf{b}_4$, we need to read values from the memory objects pointed-by $x$.  To this end, we gather abstract values stored into \textit{alloc}$_m$ that is pointed-by $x$. To do this, the sub-procedure $\texttt{ReadFromObject}$ only visits two basic blocks $\mathsf{b}_4$ and $\mathsf{b}_1$ by walking up the dominator tree.
	From $\mathsf{b}_4$, we can read the value $(\psi, \ell_4,  d)$, which is written into \textit{alloc}$_m$ at $\mathsf{b}_2$ and propagated to $\mathsf{b}_4$.
	From $\mathsf{b}_1$, we can read the value $(\psi, \ell_3,  c)$, which is written into \textit{alloc}$_m$ at $\mathsf{b}_1$ that dominates $\mathsf{b}_4$.
\end{example}

\subsection{Semi-Path-Sensitive Analysis}
\label{subsec:pathupdate}
Path-sensitivity comes in many flavors, depending on the kind of information encoded as constraints. 
Previous work on path-sensitive pointer analysis either adopts relatively coarse abstractions where Boolean variables abstract just the control flow, ignoring the actual predicate of a condition~\cite{sui2011spas}; 
or takes expensive abstractions with first-order theory formulas tracking data predicates, which can incur huge overhead~\cite{livshits2003tracking,hackett2006aliasing,dillig2011precise}.\footnote{\citet{livshits2003tracking} invoke a computer algebra system. ~\citet{hackett2006aliasing} implement a procedure similar to bit-blasting that translates arithmetic constraints to SAT constraints. \citet{dillig2011precise} use the Mistral SMT solver.}

For constructing the guarded value-flow graph, we explore a sweet spot in the space; it is \emph{semi-path-sensitive} from the two aspects below.
\begin{itemize}
	\item First, the guards are conceptually first-order formulas, but abstracted as {Boolean skeletons}. 
	For instance, we abstract the two branch conditions $x > 2$ and $x \leq 2$ to fresh Boolean literals $p$ and $\neg p$, respectively. 
	Such encoding allows for certain degrees of branch correlation tracking. The design is similar in spirit to the DPLL($T$) lazy SMT solving architecture, which separates propositional reasoning and theory reasoning in a modular and demand-driven way~\cite{DPLLT-JACM06}.
	\item Second,  instead of applying a full-featured SAT/SMT solver, we adopt several linear time semi-decision procedures  such as unit-propagation~\cite{zhang1996cient} for identifying ``easy'' unsatisfiable or valid constraints, as well as performing logical simplifications. 
	In our experiment, we observe that about 70\% of the path conditions constructed in the analysis are satisfiable. 
	For the remaining ones, 80\% of them are easy constraints and can be solved with the semi-decision procedures.
\end{itemize}

Many infeasible path-sensitive data-flow facts can be filtered because programmers tend to maintain an implicit and simple correlation of conditional points-to relations, both for ensuring some required logical properties and the good human readability. 
The semi-path-sensitive analysis makes this correlation explicit. 

\paragraph{\textbf{Benefits}}
Compared with the state of the arts that construct the value-flow graph via a flow-insensitive points-to analysis and label the edge with a memory object~\cite{sui2016sparse,yan2018spatio}, our analysis has two major benefits. 
First, 	it catches the path correlations between different statements, thus pruning more infeasible value flows than path-insensitive analyses.
Second, it concisely merges and simplifies the guards qualifying a value flow, when merging the value-flow edges between load and store statements.

\begin{figure}[t]
	\centering
	\includegraphics[width=0.36\textwidth]{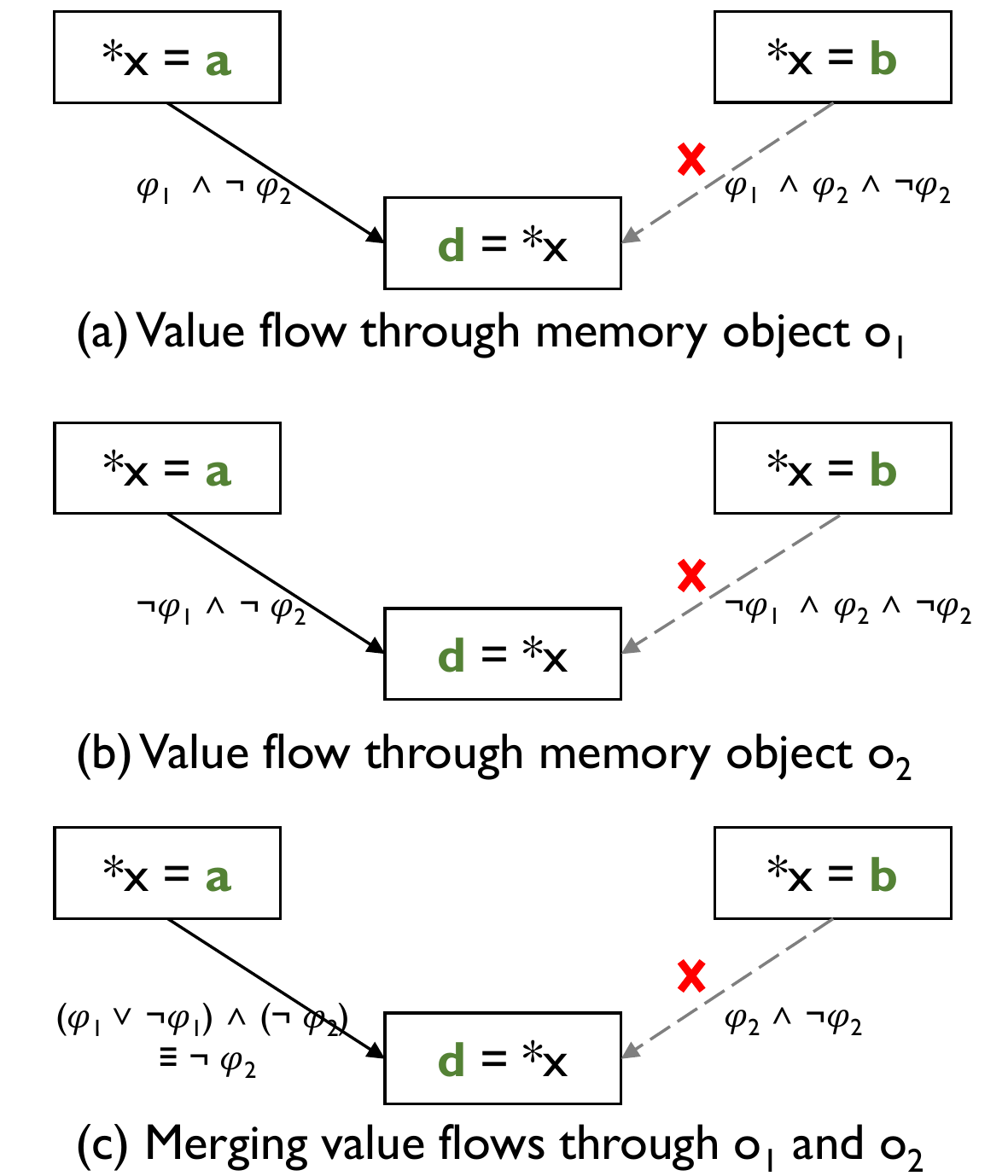}
	\caption{
		Pruning and merging value-flow graph edges for the program in Fig.~\ref{demand-merge}(a).}
	\label{update2}
\end{figure}

\begin{example}
	
Consider the program in Fig.~\ref{demand-merge}(a). 
Observe that the variable $x$ may point-to $o_1$ or $o_2$. 
A path-insensitive algorithm will conclude that $d$ may alias with $\{a, b \}$, where $b$ is a false positive. 
We now explain intuitively how our algorithm works and prunes the false positive. 
Let us consider the two cases where $x$ points-to $o_1$  or $o_2$, respectively. 
First, if $x$ points-to $o_1$, as in Fig.~\ref{update2}(a), our analysis will obtain the following values that may flow to   $d \smallletter{@} d = *x$:
\begin{enumerate}
	\item $(\varphi_1 \land \neg \varphi_2, a)$
	\item $(\varphi_1 \land \varphi_2 \land \neg \varphi_2, b)$
\end{enumerate}	
The semi-path-sensitive analysis can decide that the guard of the second item is unsatisfiable. Hence, the value $b$ is pruned.
Second, if $x$ points-to $o_2$, as in Fig.~\ref{update2}(b), the analysis can also prune the value $b$. Finally,  after merging the two  graphs induced by $o_1$ and $o_2$, we obtain the graph in Fig.~\ref{update2}(c). 
\end{example}

\paragraph{\textbf{Summary}}
The on-the-fly sparse analysis and the semi-path-sensitive analysis conspire to address the challenges of conservative propagation and constraints explosion (\cref{subsec:rules}). 
The sparse analysis skips unnecessary control flows when propagating data-flow facts, thereby improving the analysis efficiency.
The semi-path-sensitive analysis removes a lot of false pointer information, which not only improves precision but also benefits efficiency because smaller points-to sets lead to less work~\cite{lhotak2008evaluating,smaragdakis2014introspective}.

\section{Interprocedural Analysis}
\label{sec:inter}
The tenet of our interprocedural analysis is breaking down the entire abstraction into smaller components to enable on-demand resolution of alias relations. 
To achieve context-sensitivity but avoid expensive summary cloning, we first introduce the approach to building concise function summaries.
We then sketch the process of constructing inter-procedural value-flow graph.


\paragraph{\textbf{Call Graph}} A pointer analysis often faces the ``chicken-and-egg'' problem: performing the analysis requires a call graph, which also needs reasoning about function pointers~\cite{grove2001framework}.  
In our approach, we consult a flow- and context-insensitive analysis~\cite{zhang2013fast} for obtaining a sound call graph. 
Previous works~\cite{milanova2004precise,hardekopf2011flow} have shown that a precise call graph for C-like programs can be constructed using only flow-insensitive analysis. 
For C++ programs, we adopt the class hierarchy analysis to resolve virtual function calls~\cite{dean1995optimization}.

\subsection{Cloning-Based Analysis with Concise Function Summaries}
\label{subsec:inter-summary}
\begin{figure*}[t]
	\centering

	\includegraphics[width=0.73\textwidth]{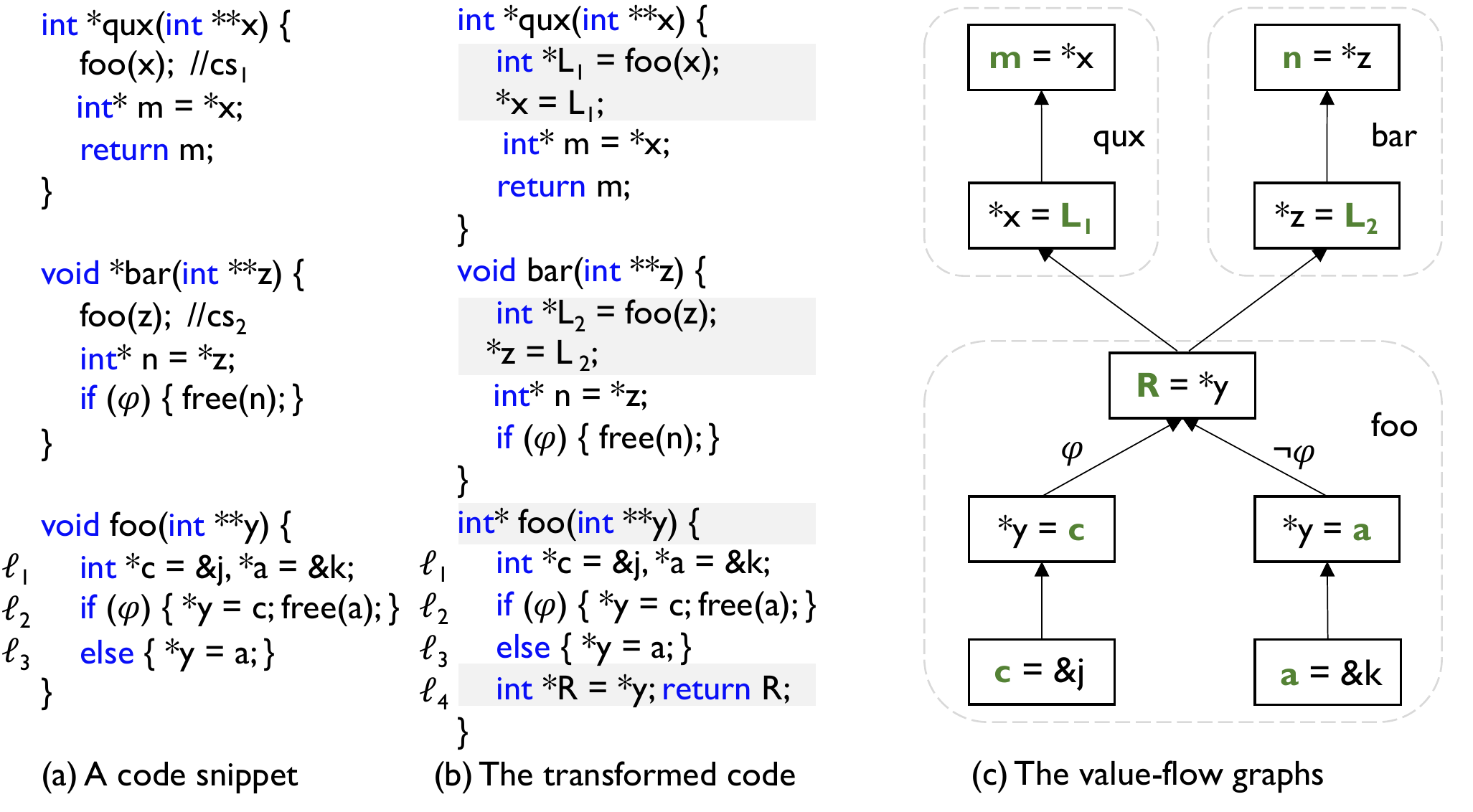}

	\caption{
		An example of using concise summary and performing on-demand search.}
	\label{ptgraph-clone}
\end{figure*}

For building interprocedural value-flow graphs, 
we perform a bottom-up and summary-based analysis.
To achieve context-sensitivity, conventional summary-based analyses
conservatively identify the side effects of a function,
which are then cloned at every call site of the summarized function in the upper-level callers~\cite{xie2005scalable,dillig2011precise}.  
However, the size of the side-effect summary can quickly explode, becoming a significant obstacle to scalability. To illustrate, consider the program in Fig.~\ref{ptgraph-clone}(a).
In conventional approaches, the summary of \textit{foo} is the points-to information, $\mathbb{E}$ and $\mathbb{S}$, of the interface variable $y$ at the exit point of \textit{foo}. That is,
\begin{align*}
\begin{split}
\mathbb{E}(y)  &=  \{ (\texttt{true}, o) \}; \ \ \mathbb{S}(o)   = \{ (\varphi, \ell_2, c),  (\neg \varphi, \ell_3, a) \}; 
\end{split}
\end{align*}

By cloning the summary to the two call sites in \textit{qux} and \textit{bar},
the two variables $a$ and $c$ are cloned twice.
When the summaries of \textit{qux} and \textit{bar} are cloned to their upper-level callers, $a$ and $c$ will continue to be cloned. 
As a result, the summary size will grow exponentially.

To mitigate the problem, our basic idea is to introduce symbolic auxiliary variables, each of which stands for a class of variables to clone.
Then we can only clone a single auxiliary variable during interprocedural analysis, reducing the burden of cloning.
For the above example, we introduce an extra value $R$ for the function \textit{foo} to represent all values (e.g., $a$ and $c$) stored in the memory object pointed-by $y$.
As a result, 
the function summary gets smaller as the following, and we only need to clone a single variable $R$ to the callers during the interprocedural analysis. 
\begin{align*}
\begin{split}
\mathbb{E}(y)  &=  \{ (\texttt{true}, o) \}; \ \ \mathbb{S}(o)  = \{ (\texttt{true}, \ell_4, R) \};  \\
R               &\mapsto \{ (\varphi, \ell_2, c),  (\neg \varphi, \ell_3, a) \} \\ 
\end{split}
\end{align*}

Intuitively,
this process amounts to adding an extra return value to the function \textit{foo}, as shown in Fig.~\ref{ptgraph-clone}(b). 
Formally, this summarization process is illustrated in Alg.~\ref{aux-summary2}, where the points-to results are always merged into a single auxiliary variable so that the burden of the cloning processes can be reduced. 
Each auxiliary variable stands for a modified non-local memory object accessed through an access path rooted at an interface variable.
In conclusion,
the summarization scheme enables local reasoning of the value flows, as opposed to global reasoning about the entire heap. 

\begin{algorithm}[t]
	\caption{Summarize an interface variable}
	\label{aux-summary2}
	\KwIn{An interface variable (parameter or return) $x$}
	\KwOut{Updated $\mathbb{E}, \mathbb{S}$ with auxiliary variables}
   $WL \leftarrow \{ x\}$\;
   \While{$WL$ is not empty} {
   	$x \leftarrow$ pop an element from $WL$\;
   	$R, o_x \leftarrow $ create an auxiliary variable and a memory object\;
   	\For{$(\pi, o) \in  \mathbb{E}(x)$} {	
   		\For{$(\varphi, \ell', v) \in \Pi_{\pi}(\mathbb{S}(o))$}{	
   			$\mathbb{E}(R) \leftarrow \mathbb{E}(R) \cup \Pi_{\varphi}(\mathbb{E}(v))$ \;	
   		}
   	}
    $\mathbb{E}(x) \leftarrow \{ (\texttt{true}, o_x) \}$,
    $\mathbb{S}(o_x) \leftarrow \{ (\texttt{true}, \_, R) \}$\;
   	\If{$R$ is a pointer}{ $WL \leftarrow WL \cup \{ R\}$}
   }   
\end{algorithm}

\subsection{Constructing the Value-Flow Graphs}
\label{subsec:inter-query}

Alongside the analysis, we have been able to construct the value-flow graph (\cref{sec:domain}) for each function. 
The graph has two types of edges representing value-flow relations: the \emph{direct edge} connects a store to a load  using the rule in Fig.~\ref{vfrule},  and the \emph{summary edge} connects $s$ to $t$ if 
$s$ can transitively flow to $t$. 
The analysis eagerly connects the summary edges 
between a formal argument (formal-in) at the function entry and the return value (formal-out) at the function
exit.  
As shown in Fig.~\ref{ptgraph-clone}(c),
the local value-flow graphs are stitched together by matching formal and actual parameters as
well as return value and its receivers.




\section{Answering Demand Queries}

The storeless value-flow graphs allow tracking of transitive data dependence in a demand-driven fashion. 
By a forward or backward graph traversal, the graphs can be adopted in various applications.

\paragraph{\textbf{Thin Slicing for Program Understanding}}
The first typical application is \textit{thin Slicing}~\cite{sridharan2007thin,li2016program}, which can be implemented via a backward traversal on the value-flow graph.
Thin slicing is introduced by~\citet{sridharan2007thin} to facilitate program debugging and understanding. 
A thin slice for a program variable, a.k.a., the slicing seed, includes only the \emph{producer statements} that directly affect the values of the variable.
In contrast to conventional slicing, control dependence and the data dependence of the variable's base pointer are excluded. Hence, thin slices are typically much smaller than conventional program slices. 

\begin{example}
Consider the program in Figure~\ref{ptgraph-clone}. 
To build the thin slice for the slicing seed $m$ at function \textit{qux}, we only need to traverse the value-flow graphs from $m$ in a reversed direction.
Such a traversal will visit all program statements that need to be included in the slice. For instance, $*y = a$ and $*y = b$ will be visited and included in the thin slice as they are the producer statements of the variable $m$.
\end{example}

%
%
%

\paragraph{\textbf{Value Flow Analysis for Bug Hunting}}
The second typical application is \emph{value flow analysis}~\cite{fastcheck-07,sui2012static}.
The analysis of value flows
underpins the inspection of a broad range of software bugs,
such as the violations of memory safety (e.g., null dereference, double free, etc.),
the violations of resource usage (e.g., memory leak, socket leak, etc.),
and security problems (e.g., the use of tainted data).
Clearly, it is of vital importance to precisely resolve value flows caused by pointer aliasing, which is the key problem we address in the paper. 
Value flow analysis can be implemented via a forward traversal on the value-flow graphs, during which the alias constraints and property-specific constraints can be gathered together and handed to an SMT solver.


\begin{example}
Suppose we need to detect double-free bugs for the program in Figure~\ref{ptgraph-clone}(a). 
We traverse its value-flow graphs (Figure~\ref{ptgraph-clone}(c)) starting  from $a$ and obtain one path from $a$ to $n$. We then stitch together the 
path condition under which $n$ is data-dependent on $a$ (i.e., $\neg \varphi$),  and the path conditions of the two 
statements \textit{free(a)} and \textit{free(n)} (i.e., $\varphi \land \varphi$). 
Observe that we do not compute the interprocedural data dependence between $(a, m)$ or $(c, n)$.
\end{example}

When used for full path-sensitive value flow analysis,
our design has two advantages.
First, it enhances precision, as a combined domain of pointer and source-sink information allows more precise information than could be obtained by solving each domain separately~\cite{Cousot:1979:SDP:567752.567778,fink:typestate:issta}. 
Second, it allows scalability, because the client (1) does not need to perform explicit cast-splitting  over the points-to sets when handling indirect reads/writes, 
alleviating a major source of case explosion in previous work~\cite{PSE,blackshear2013thresher,yan2018spatio};
(2) sparsely tracks the value flows by following the data-dependence edges;
and (3) can make distinctions between memory objects summarized by an access path on demand.
Consequently, the client concentrates computational effort on  the path- and context-sensitive pointer information only when it matters to the properties of interest.	

\paragraph{\textbf{Summary}}
Both the value-flow graphs building phase and the on-demand analysis phase are sparse, by piggybacking the computation of pointer information with the resolution of data dependence. 
The two phases mitigate the aliasing-path-explosion problem as follows.
First, when constructing the value-flow graphs, duplicate edges are merged, many false edges are pruned, and the data-dependence guards are significantly simplified (\cref{sec:intra}).
Second, when resolving transitive data dependence over the graphs,
the concise summaries serve as  the ``conduits'' to allow values of interests flow in and out of the function scope.
In particular, the client
 (1) does not need to perform explicit cast-splitting  over the points-to sets when handling indirect reads/writes, 
alleviating a major source of case explosion in previous work~\cite{PSE,blackshear2013thresher,yan2018spatio};
(2) can sparsely track the value flows by following the data-dependence edges;
and (3) can make distinctions between memory objects summarized by an access path on demand.
Consequently, the client analysis can concentrate computational effort on  the path- and context-sensitive pointer information only when it matters to the properties of interest.

\section{Evaluation}
\label{sec:evaluation}

To demonstrate the utility of \ToolName, we examine its scalability in constructing the value-flow graphs (\cref{subsec:svdg}), and apply it to two practical clients, namely semi-path-sensitive  thing slicing (\cref{subsec:eval-pts-slice}), and fully path-sensitive  bug hunting (\cref{subsec:alias-pair}).

\paragraph{\textbf{Implementation}}
We have implement \ToolName\ on top of 
LLVM 3.6. While the language in ~\cref{sec:domain} has restricted language constructs, our implementations support most features of C/C++, such as unions, arrays, classes, dynamic memory allocation, and virtual functions. 
Arrays are considered monolithic.
\ToolName\ is \emph{soundy}~\cite{livshits2015defense}, which means that it handles most language features in a sound manner, while it also applies some unsound choices as in previous works ~\cite{babic2008calysto,xie2005scalable}.  
For instance, we unroll each loop twice on the control flow graph and call graph, do not handle inline assembly, assume distinct parameters are not alias with each other.

\subsection{Experimental Setup}

\paragraph{\textbf{Baselines}}
In this section,
we compare \ToolName\ against three groups of analyses.
\begin{enumerate}
	\item First, we compare with the following analyses for constructing the value-flow graphs: (1) \svf~\cite{sui2016svf}, an inclusion-based, flow- and context-insensitive pointer analysis; \footnote{\url{https://github.com/SVF-tools/SVF}} 
(2) \sfs~\cite{hardekopf2011flow}, an inclusion-based, flow-sensitive, context-insensitive pointer analysis;
(3) \dsa~\cite{lattner2007making}, a unification-based, flow-insensitive, context-sensitive pointer analysis. \footnote{\url{https://github.com/seahorn/sea-dsa}} 
    \item Second, for the thin slicing client, we compare with \supa~\cite{sui2016demand,sui2018vfdemand}, \footnote{\url{https://github.com/SVF-tools/SUPA}}
the state-of-the-art demand-driven flow- and context-sensitive pointer analysis for C/C++. 
\supa\ relies on \svf\ to build the value-flow graphs, base on which it answers demand queries. 
   \item Finally, we compare with \cred~\cite{yan2018spatio}, a state-of-the-art path-sensitive pointer analysis for bug hunting.\footnote{The tool is not open-source. We implement the algorithm on top of \svf.}
\end{enumerate}
All of these analyses are filed-sensitive, meaning that each field of a struct is treated as a separate variable.

We cannot compare with the pointer analyses in~\cite{guyer2005error,yu2010level,li2011boosting,li2013precise,sui2011spas,sui2014making,zhao2018parallel} because their implementations are not publicly available. 
For bug finding, we tried our best to compare with \saturn~\cite{hackett2006aliasing} and \compass~\cite{dillig2010fluid,dillig2011precise}, but they are not runnable on the experimental environment that we are able to set up.


\paragraph{\textbf{Subjects and Environment}}
Tbl.~\ref{eval:pts-customized} shows the benchmarks.
Six of them are taken from SPEC CINT2000 and ten from open-source
projects.
The programs cover a wide range of applications such as text editors and database engines, and their sizes vary from 13 KLoC to 8 MLoC. Note that
since \ToolName\  unrolls loops on the control flow graph and call graph, we feed the same transformed code to other tools. 
All experiments are conducted on a 64-bit machine with 40 Intel Xeon E5-2698 CPUs@2.20 GHz and 256 GB of RAM. 
All runtime numbers are medians of three runs.

\subsection{Value-flow Graph Construction}

\label{subsec:svdg}
First, we examine the scalability of \ToolName\ for constructing value-flow graphs. 
The cutoff time per tool  per program is 12 hours.

\begin{figure}[t]
	\centering
	\includegraphics[width=0.35\textwidth]
	{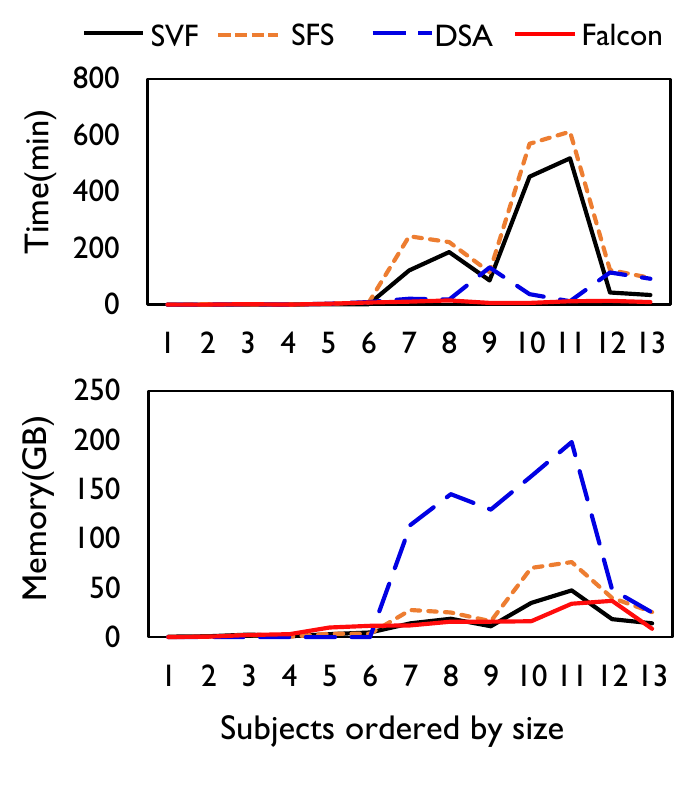}
	\caption{
		Comparing time and memory cost of 
		value-flow graphs construction.}
	\label{vfgtime-fig}
\end{figure}

\paragraph{\textbf{Comparing with \svf}, \textbf{\sfs}, \textbf{and \dsa}} Tbl.~\ref{eval:pts-customized} and Fig.~\ref{vfgtime-fig} 
show the results of the four analyses. In terms of runtime overhead, 
we can see that they perform similarly in small-sized programs. However, on programs with more than 500 KLoC, \svf\ and \sfs\ get derailed and become orders-of-magnit\-ude more expensive. 
In particular, both fail to analyze \texttt{mysql}, \texttt{rethinkdb}, and \texttt{firefox} within 12 hours. 
\dsa\ is comparable to \ToolName\ on  \texttt{vim} and \texttt{php}, but much slower than \ToolName\ on other large programs (\texttt{git}, \texttt{wrk}, \texttt{libicu}, and \texttt{mysql}). 
Also, \dsa\ cannot finish the analysis of \texttt{rethinkdb} and \texttt{firefox}. 
To sum up, \ToolName\ is on average 17$\times$, $25\times$, and 4.4$\times$  faster than \svf, \sfs, and \dsa, respectively. 
In terms of memory consumption, on average, \ToolName\ takes  1.4$\times$, 1.9$\times$, and 4.2$\times$ less memory than \svf, \sfs, and \dsa, respectively.

\begin{table}[t]
	\caption{Benchmark size (\textbf{KLoC}) and runtime (in \textbf{minutes}) of building value-flow graphs.}
	\label{eval:pts-customized}
  \resizebox{0.48\textwidth}{!}
    {
	\begin{tabular}{|l|c|c|c|c|c|c|c|}
		\hline
		\textbf{Program} & \textbf{Size} &\textbf{\svf} & \textbf{\sfs} & \textbf{\dsa} & \textbf{\ToolName} & \textbf{\ToolName(\textsc{Pi})} & \textbf{\ToolName(\textsc{Sat})}\\ \hline
		crafty  & 13          &  $<$0.1  & $<$0.1  &  $<$0.1      & $<$0.1  & 0.2 & 1.9    \\ 
		eon     & 22          & 0.3      & 0.7  & 0.2          &  0.3   & 0.4 & 2.1         \\ 
		gap     & 36          & 0.4      & 1.7  & 0.9          &  0.5   & 0.7 & 13       \\ 
		vortex  & 49          & 0.1      & 0.2  & $<$0.1       &  0.3   & 0.3 & 26          \\ 
		perlbmk & 73          & 0.7      & 2.7  & 2.1          &  2.1   & 3.2& 125     \\ 
		gcc     & 135         & 0.7      & 7.4  & 8.6         &  7.8   & 8.1 & 145       \\ \hline 
		git     & 185         & 121     & 243 & 21          & 10     &16 &393  \\
		vim     & 333         & 186     & 221 & 17          & 14   & 23  & 484         \\ 
		wrk     & 340         & 85      & 115 & 131         & 6   & 15 & 537      \\  
		libicu  & 537         & 454     & 570 & 37          &	5    & 6   & OOT  \\  
		php     & 863         & 519	  & 614 & 12          & 12   & 41  & OOT  \\ 
		ffmpeg  & 967         & 43	  & 122 & 113         & 13	 & 36 & OOT \\ 
		ppsspp  & 1648        & 34	  & 94  & 92          & 8	 & 25  & OOT \\ 
		mysql   & 2030        & OOT     & OOT & 113         & 46   & 64 & OOT  \\ 
		rethinkdb  &3776     & OOT     & OOT & OOT         & 90   & 113 & OOT  \\ 
		firefox & 7998        & OOT 	  & OOT & OOT         & 167	 & 273 & OOT \\ 
		\hline
		  \multicolumn{8}{l}{\scriptsize OOT means the analysis runs out of the time budget (12 hours). }
	\end{tabular}
  }
\end{table}


We attribute the graceful scalability of \ToolName\ to two factors. First, the combination of on-the-fly sparsity and semi-path-sensitivity translates into significant gains in both precision and performance (\cref{sec:intra}).
Second, we do not clone the full points-to information to achieve context-sensitivity. Instead, we leverage a concise summary to avoid computing a whole-program image of the heap~(\cref{sec:inter}). 

\paragraph{\textbf{The Effects of Semi-Decision Procedures}}
To understand the effects of constraint solving, we set up two additional configurations of \ToolName\ for constructing value-flow graphs. 
Specifically, \ToolName-PI is path-insensitive, while  \ToolName-SAT uses a full-featured SAT solver. The last three columns of Tbl.~\ref{eval:pts-customized} compare the three configurations.
\ToolName\ is usually more--and occasionally much more--efficient than \ToolName-PI, due to the increased precision. 
However,  \ToolName-SAT is not a good choice in practice: its precision is offset by unbearable runtime overhead. 
In particular, \ToolName-SAT runs out of the time budget for all programs of more than 500 KLoC.


The results indicate that solving constraints when building value-flow graphs pays off, which naturally raises the question: could we do better by tuning the semi-decision procedure more aggressively? However, we find that being ``too aggressive'' can lead to performance overhead that overwhelms the benefits. For instance, we tried the $O(n^3)$ Gaussian elimination algorithm for solving linear constraints, leaving the analysis hard to scale to millions of lines of code. Adapting the decision procedures defines a sophisticated design space that deserves further optimizations.

\begin{figure}[t]
	\centering
	\includegraphics[width=0.45\textwidth]{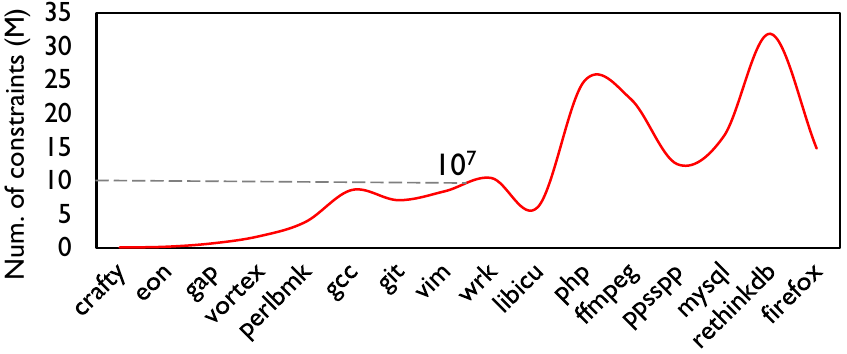}
	\caption{
		Number of constraints \ToolName\ deals with when building value-flow graphs.}
	\label{num-cnts-fig}
\end{figure}

To provide insight into the nature of constraints explosion, in Fig.~\ref{num-cnts-fig}, we report the number of constraints \ToolName\ deals with. Even when unrolling all loops, we can see that it is not unusual to have over $10^7$ constraints.


%
%
%
\subsection{ Thin Slicing for Program Understanding}
\label{subsec:eval-pts-slice}
This study aims to give a measure for the precision of  \ToolName's semi-path-sensitive value-flow graphs (\cref{subsec:pathupdate}) with the thin slicing  client. 
In the experimental results, we exclude the time for building value-flow graphs.

To generate a realistic set of
queries,
we  compare \ToolName\ against \supa\ for thin slicing, which is introduced by~\citet{sridharan2007thin} to facilitate program  understanding and debugging.
The thin slice for a given pair of  program variable and statement, a.k.a., the slicing seed, includes only the \emph{producer statements} that directly affect the values of the variable.
In contrast to conventional slicing, control dependence and the data dependence of the variable's base pointer are excluded. 
Hence, thin slices are typically much smaller than conventional program slices. 

\begin{figure}[t]
	\centering
	\includegraphics[width=0.48\textwidth]{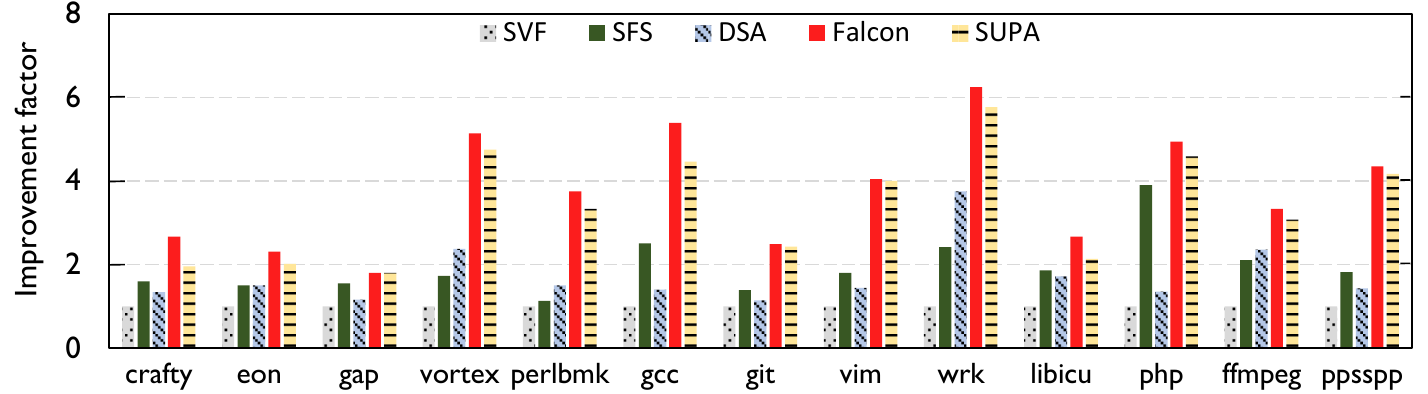}
	\caption{
		Improvement in average slice size compared with the baseline \svf.
	}
	\label{eval:pts}
\end{figure}

We generate the queries from the bug reports issued by a third-party typestate analysis, 
which only flags the buggy variables and program locations, but not the trace under which the bugs may occur.
Thus, the thin slices can assist the developers in understanding the bug reports. 
Our results show that \ToolName\ is scalable for the thin slicing client, taking under 240 milliseconds for each demand query. In summary, it achieves up to 302$\times$ speedups than \supa\ and 54$\times$ on average. 
Fig.~\ref{eval:pts} compares the precision of \ToolName\ against \, \sfs, \dsa, and \supa\ on the 13 programs that get analyzed by all tools. 
The data for each program are normalized to
the results of \svf, i.e, a higher bar corresponds to a more precise analysis.
Briefly, we make the following observations:
\begin{itemize}
	\item The precision of \ToolName\ is superior than other analyses. The average size of slices produced by \ToolName\ is 5.5$\times$, 1.9$\times$, 2.6$\times$, and 1.3$\times$ smaller than that of \svf, \sfs, \dsa, and \supa, respectively.
	\item Comparing \sfs\ and \svf, we see that flow sensitivity can substantially improve the precision of Andersen's analysis in some programs, such as  \texttt{php} and \texttt{ffmpeg}.
	\item \dsa\ is comparable to \svf\ in some cases, and much more precise than \svf\ in many programs
	(e.g., \texttt{vim}, \texttt{libicu}, \texttt{ffmpeg}).
	The combination of context-sensitivity and unification may bring better precision than the
	flow- and context-insensitive Andersen's analysis.
\end{itemize}

\paragraph{\textbf{Remarks}}
First, 
\ToolName\ offers a clearly visible improved precision of pointer information.
An important reason is that \ToolName's semi-path-sensitive analysis (\cref{subsec:pathupdate}) can prune away more spurious value flows, compared with the demand-driven flow- and context-sensitive analysis offered by \supa. 

Second, the performance improvement is because that the value-flow graphs of \ToolName\ are more compact than that of \supa. 
On the one hand,
\supa\ constructs the graphs with a flow-insensitive analysis, whereas \ToolName\ uses a flow-sensitive one.
On the other hand, \supa\ has to explicate a memory object on en edge so that it can
answer demand queries, where many edges are redundant.

Third, we observe that \ToolName's unsound assumption that function parameters are alias-free does not affect the
soundness for more than $90\%$ of the queries, validated by manually checking the results. Similar to our observations, two recent studies~\cite{sui2014making,gharat2016flow} also show that the function parameters of real-world C/C++ programs tend to have few aliasing relations.




\subsection{Value Flow Analysis for Bug Hunting}
\label{subsec:alias-pair}
In this study, we investigate the efficiency and effectiveness of \ToolName\ for double-free detection, by comparing it against \cred~\cite{yan2018spatio}. 
Both \ToolName\ and \cred\ employ the Z3 SMT solver~\cite{de2008z3} to achieve full path-sensitivity. 
Each tool is run in a single-thread mode and the cutoff time per program is 24 hours.

\begin{table}[t]
	\centering
	\caption{Results of double-free detection.}
	\label{eval:bughunt-dbf}
   \resizebox{0.48\textwidth}{!}
  {
	\begin{tabular}{ | l |c c c  |c c c  |}
		\hline
		\multirow{3}{*}{\textbf{Program}} & \multicolumn{3}{c|}{\textbf{\cred}} &  \multicolumn{3}{c|}{\textbf{\ToolName}}\\ \cline{2-7} &  %
		Time(m) &  Mem(G) &   \#FP/\#Rep  & Time(m) &  Mem(G) &   \#FP/\#Rep   \\ 
		\hline
		crafty        & 0.2    & 0.5       & 0/0      & 0.3       & 0.5     &0/0   \\ 
		eon           & 1.3    & 1.1      & 0/0      & 0.9         & 1.7      &0/0    \\ 
		gap           & 0.7    & 3.2      & 0/0      & 0.6       & 1.3      &0/0    \\ 
		vortex        & 0.4    & 1.4      & 1/1      & 0.5       & 4.3      &0/0    \\ 
		perlbmk       & 5.2    & 4.4      & 0/0      & 3.4       & 4.8      &0/0  \\ 
		gcc           & 16.1   & 5.7      & 2/2      & 9.4       & 1.3      &0/0    \\ \hline		               
		git           & 155     & 23       & 1/1     & 23       & 20       &1/3   \\ 
		vim           & 1100   & 63       & 0/0     & 112      & 27        &0/0    \\ 
		wrk           & 93     & 19       &0/0      & 17       & 24      &0/0    \\ 
		libicu        & 824    & 78       &0/0     & 79       & 28       &0/2  \\
		php           & 1310    & 84       &5/11   & 26      & 51      &2/9   \\ 
		ffmpeg        & 56     & 61       &2/7    & 21      & 54       &2/8    \\ 
		ppsspp        & 44     & 33       &0/1   & 23       & 17        &1/3   \\ 
		mysql         & OOT    & -        & -     & 178      & 74        &1/3    \\ 
		rethinkdb     & OOT    & -        & -      & 274      & 108       &0/0    \\ 
		firefox       & OOT    & -        & -      & 310      & 143         &0/0   \\ 
		\hline 
		\textbf{\%FP} & \multicolumn{3}{c|}{47.8\% (11/23)} & \multicolumn{3}{c|}{25\% (7/28)} \\
		\hline  
		\multicolumn{7}{l}{OOT means the tool runs out of the time budget (24 hours).}
	\end{tabular} 
  }
\end{table}

Tbl.~\ref{eval:bughunt-dbf} presents the time and memory overhead of the tools.
As can be seen, 
\ToolName\ surpass the performance of \cred\ for most large-scale programs, achieving up to 50$\times$ speedups, and 6$\times$ speedups on average. 
Besides, on average,  \ToolName\ takes 2$\times$ less of memory than \cred.
Although not shown in the table, we remark that, if allowing the concurrent analysis of 10 threads, \ToolName\ can finish the checking of each program
within 45 minutes.
Tbl.~\ref{eval:bughunt-dbf} also shows the number of reported warnings (\#Rep) as well as the number and the rate of false positives (\#FP and \%FP). 
We can see that \ToolName\ detects more real vulnerabilities than \cred\ (21 vs. 12). 
The false-positive rates of \ToolName\ and \cred\  are 25\% and 56.5\%, respectively.
\ToolName\ is aligned with the common industrial requirement of 30\% false positives~\cite{Bessey:2010:FBL:1646353.1646374,McPeak:2013:SIS:2491411.2501854}.

Overall, our findings conclude that the ideas behind \ToolName\ have considerably practical value. In terms of all aspects, including not only the scaling efforts but also the precision and recall, the tool itself is promising in providing an industrial-strength capability of static bug hunting.

\section{Related Work}
\label{sec:related}
 \paragraph{ \textbf{Path-Sensitive Analysis}}
Tbl.~\ref{related-table}  gives key properties of several existing path-sensitive algorithms.
Here we summarize some of the approaches taken by previous analyses.
 with a focus on pointer reasoning.
\begin{table}[t]
	\caption{A comparison of key properties of existing analyses.
		The ``Inter-PS'' column represent interprocedurally path-sensitive. The ``PS-Heap'' and `Sparse'' columns respectively indicate whether the analysis uses path-sensitive heap abstraction and is sparse. Finally, the ``Shown to Scale'' column indicates whether the algorithm has been shown to scale to large  programs with multi-million lines of code.
		}
	\label{related-table}
   \resizebox{0.48\textwidth}{!}
   {
		\begin{tabular}{| l | c | c | c | c |}
			\hline
			\textbf{Algorithm }& \textbf{Inter-PS}  &\textbf{PS-Heap} & \textbf{Sparse}& \begin{tabular}[c]{@{}l@{}}\textbf{Shown to Scale}\end{tabular} \\ \hline
			\citet{das2002esp}                     &  X     &    &        &             X         \\ 
			\citet{ball2002s}                         &  X      &   &        &                       \\ 
			\citet{livshits2003tracking}        &  X     &X &        &                     \\ 
			\citet{hackett2006aliasing}        &        & X &        &       X           \\ 
			\citet{babic2008calysto}            &  X      & &        &        X          \\ 
			\citet{Snugglebug}                       &  X    & X  &        &                  \\ 
			\citet{dillig2011precise}               & X      &X&       &                     \\ 
			\citet{sui2011spas}                      &    & X &  X      &       X                      \\ 
			\citet{blackshear2013thresher}  & X     & X &        &                             \\ 
			\citet{li2017semantic}                 &   X   & X&        &                 \\ 
			\citet{yan2018spatio}                  &   X     & X   & X       &                  \\ 
			\citet{Focal-FSE19}                     &   X     &X&         &                      \\ 
			Current paper                              &   X     &X&  X      &    X                          \\ \hline
		\end{tabular}
	}
\end{table}
\citet{livshits2003tracking} introduce a flow-, path-, and context-sensitive pointer analysis, which only scales to programs up to 13KLoC.
The pointer analyses in~\cite{hackett2006aliasing} and  ~\cite{sui2011spas} are only intraprocedurally path-sensitive.
\citet{dillig2010fluid,dillig2011precise} present a path- and context-sensitive heap analysis that scales to program with 128KLoC.
~\citet{blackshear2013thresher} introduce a symbolic-explicit representation that incorporates the pre-computed flow-insensitive points-to facts to guide the backward symbolic execution. 
Similar to the index variables in~\cite{dillig2010fluid,dillig2011precise} and symbolic variables in~\cite{blackshear2013thresher}, we use the guards qualifying value-flow graph edges to enable lazy case splitting over the points-to set. 
However, their approaches are either not demand-driven or non-sparse and, thus, do not scale well for large-scale programs.

Our work follows a long line of research on path-sensitive dataflow analysis. 
\textsc{Esp}~\cite{das2002esp} encodes a typestate property into a finite state automata, which is used as criteria for partitioning and merging program paths.
Essentially, \textsc{Esp} is similar to many other approaches such as \emph{trace partition}~\cite{mauborgne2005trace} and \emph{elaborations}~\cite{sankaranarayanan2006static} that control the trade-off between performing joining operations or logical disjunctions at control flow merge points.  
By contrast, \ToolName\ uses logical disjunction to precisely merge value-flow guard.

Counterexample-guided abstraction refinement starts with an imprecise abstraction, which is iteratively and gradually refined~\cite{CEGAR-JACM,ball2002s}. 
Our approach has a ``refinement'' flavor, but we compute the path- and context-sensitive conditions directly in the on-demand analysis phase, without using a refinement loop. 
Conceptually, our approach bears similarities to the staged analyses for typestate verification \cite{fink:typestate:issta,fink2008effective}, but we focus on path-sensitive analysis and 
We decompose the cost into semi-path-sensitive value-flow graphs construction and a full-path-sensitive alias resolution over the graphs.

Shape analysis~\cite{TVLA-TOPLAS} proves data-structure invariants and has had a major impact on the verification community. 
Precise shape analyses~\cite{TVLA-TOPLAS,li2017semantic} that are capable of path-sensitive heap reasoning do not readily scale to large programs~\cite{fink2008effective}. 
There have been scalable solutions such as compositional shape analysis based on bi-abduction ~\cite{calcagno2011compositional,gulavani2009bottom}, yet they do not guarantee precision.


\paragraph{\textbf{Demand-Driven Pointer Analysis}}
Demand-driven program analyses only analyze parts of the program that are relevant for
answering a given query.
To date, 
most existing demand-driven pointer analyses for C/C++~\cite{heintze2001demand,saha2005incremental,zheng2008demand} and Java ~\cite{sridharan2005demand,sridharan2006refinement,yan2011demand,shang2012demand,lu2013incremental,Su2014Parallel,feng2015explorer} are flow-insensitive. 
Their underlying data structures, such as the pointer expression graph~\cite{zheng2008demand}, entirely or partially lose the control flow information and, thus, are 
not easy to be extended for path-sensitivity. 
Recently, there has been a resurgence of interest in demand-driven flow- or path-sensitive pointer analysis~\cite{spath2016boomerang,sui2016demand,spath2019context,yan2018spatio}. 
Some of these approaches are not sparse ~\cite{spath2016boomerang,spath2019context}. 
Some of them are sparse but suffer from the aliasing-path-explosion issues that we address in this paper~\cite{sui2016demand,yan2018spatio}.
 
 
There is an increasing interest in \emph{parametric pointer analyses} ~\cite{kastrinis2013hybrid,smaragdakis2014introspective,wei2015adaptive,jeong2017data,jeon2018precise,hassanshahi2017efficient,li2018scalability,li2018precision} that resemble demand-driven approaches. 
By contrast, they are not query-driven, but schedule analysis strategies such as selective context-sensitivity for different parts of the program. 
(functions, allocation sites) 
\emph{Introspective analysis}~\cite{smaragdakis2014introspective} tunes context sensitivity per-function based on a pre-analysis that computes heuristics such as ``total points-to information''.
~\citet{jeong2017data} present a data-driven approach to guiding selective context-sensitive analysis, which assigns each function a context length based on a set of program features. 
Most of the recent advances focus on context-sensitive analysis.
Our approach uses a flow-insensitive analysis for  function pointers, 
and precise path- and context-sensitive analysis for other pointers of interest. 




\paragraph{\textbf{Data-Dependence Analysis}}
Data-dependence analysis aims to identify the def-use information in a
program. It has many applications such as change-impact analysis~\cite{arnold1996software,Orso:2004:ECD:998675.999453,acharya2011practical}, program slicing~\cite{sridharan2007thin,li2016program} and bug hunting~\cite{tripp2009taj,tripp2013andromeda,arzt2014flowdroid}. 

There is a huge amount of literature on context-sensitive data-dependence analysis via context-free language (CFL) reachability ~\cite{spath2019context,horwitz1990SDG,Chatterjee:2017:ODR:3177123.3158118,reps1998program2,sridharan2006refinement}, or other language reachability problems ~\cite{Tang:POPL2015,Zhang:LCL}. 
\citet{Tang:POPL2015} propose a TAL-reachability formulation that improves the scalability of on constructing function summaries of library code.  
\citet{Zhang:LCL} introduce linear conjunctive language (LCL) reachability for approximating the interleaved matching-parentheses problem of filed- and context-sensitive data-dependence analysis. 
\citet{spath2019context} present a context- and flow-sensitive data-flow analysis based on synchronized pushdown systems.
\textsc{P/taint}~\cite{PTaint-OOPSLA} unifies points-to  and taint analysis by extending the Datalog rules of the underlying points-to analysis and then computing the information all together.
All of these techniques are path-insensitive. 

The array dependence analysis~\cite{maydan1991efficient} community has developed several path-sensitive approaches that are based on SMT solving~\cite{mohammadi2018extending}, quantifier elimination~\cite{Mohammadi:2019}, among others~\cite{pothen2004elimination}. 
Typically, these approaches focus on array manipulating programs and do not scale to large-scale software with complicated pointer operations.


\paragraph{\textbf{Sparse Pointer Analysis}}
The idea of sparse analysis starts from the static single assignment (SSA) form where def-use chains are explicitly encoded~\cite{cytron1989efficient,choi1991automatic,chow1996effective}.
Such def-use chains allow the propagation of data-flow facts to skip unnecessary control flows. \citet{hardekopf2009semi} present an inclusion-based and {semi-sparse} flow-sensitive pointer analysis by leveraging the partial SSA form  in LLVM~\cite{lattner2004llvm}. It is semi-sparse because it only utilizes the def-use chains of the top-level pointers. 

To be fully sparse, the def-use information of other address-taken variables is needed.
There are two classes of full-sparse analysis.
First, the staged approaches~\cite{hardekopf2011flow,sui2016sparse, sui2016demand,yan2018spatio} exploit a lightweight and imprecise pre-computed pointer analysis to approximate the def-use relations.
Owing to the imprecision, spurious value flows will be introduced and harm the performance. Second, the on-the-fly approaches~\cite{chase1990analysis, yu2010level,sui2011spas,li2011boosting} construct the def-use chains alongside the pointer analysis. However, their approaches are exhaustive and, thus, do not scale well when path-sensitivity is required. 
Specifically, \textsc{Spas}~\cite{sui2011spas} is the only previous pointer analysis that is both path-sensitive and on-the-fly sparse, 
but they achieve incremental sparsity by extending the level-by-level analysis~\cite{yu2010level}, which must be exhaustive because
(1) a single level can consist of pointers from the whole program, 
and (2) pointers with higher levels strictly depend on the analysis results of pointers with lower levels.

\section{Conclusion}
\label{sec:conclusion}
We have presented \ToolName, our approach to path-sensitive data-dependence analysis. 
At its heart stands an analysis that concisely constructs the guarded and compositional value-flow graphs, which allow tracking path- and context-sensitive def-use information on demand.
The graceful scalability and high precision of \ToolName\ rest on our solution to the aliasing-path-explosion problem.
Specifically, \ToolName\ is sparse, demand-driven, and can prune and simplify path constraints at an early stage of the analysis. 
Our work presents strong evidence that path-sensitive data-dependence analysis is a reasonable choice for millions of lines of code. 

\balance
\bibliography{main}

\end{document}